\documentclass[aps,prl,reprint,groupedaddress,showpacs,floatfix]{revtex4-1}

\usepackage{graphicx}
\usepackage{dcolumn}
\usepackage{bm}
\usepackage[mathlines]{lineno}

\begin{document}

\title{High stability of faceted nanotubes and fullerenes of \\
       multi-phase layered phosphorus:
       A computational study}

\author{Jie Guan}
\affiliation{Physics and Astronomy Department,
             Michigan State University,
             East Lansing, Michigan 48824, USA}

\author{Zhen Zhu}
\affiliation{Physics and Astronomy Department,
             Michigan State University,
             East Lansing, Michigan 48824, USA}

\author{David Tom\'{a}nek}
\email%
{tomanek@pa.msu.edu}%
\affiliation{Physics and Astronomy Department,
             Michigan State University,
             East Lansing, Michigan 48824, USA}

\date{\today} 

\begin{abstract}
We present a paradigm in constructing very stable, faceted
nanotube and fullerene structures by laterally joining nanoribbons
or patches of different planar phosphorene phases. Our {\em ab
initio} density functional calculations indicate that these phases
may form very stable, non-planar joints. Unlike fullerenes and
nanotubes obtained by deforming a single-phase planar monolayer at
substantial energy penalty, we find faceted fullerenes and
nanotubes to be nearly as stable as the planar single-phase
monolayers. The resulting rich variety of polymorphs allows to
tune the electronic properties of phosphorene nanotubes (PNTs) and
fullerenes not only by the chiral index, but also by the
combination of different phosphorene phases. In selected PNTs, a
metal-insulator transition may be induced by strain or changing
the number of walls.
\end{abstract}




\maketitle

One reason for the unprecedented interest in graphitic carbon is
its ability to form not only self-supporting graphene
layers~\cite{{Novoselov04},{Zhang05}}, but also single- and
multi-wall nanotubes~\cite{Iijima91} and
fullerenes~\cite{Kroto85}. Similar to graphite, which is the
parent compound of these carbon allotropes, the stable black
phosphorus allotrope is a layered compound that can be exfoliated
to phosphorene monolayers~\cite{{DT229},{Li2014}}. Phosphorus
nanotubes~\cite{{blackpnt},{Seifert2000}} and
fullerenes~\cite{{karttunen08},{Han04},{Seifert01}} have been
postulated to form in analogy to their carbon counterparts by
deforming a phosphorene monolayer, typically at significant energy
cost. In contrast to the unique structure of planar graphene, at
least four equally stable phases with different properties,
$\alpha$-P, $\beta$-P, $\gamma$-P and $\delta$-P, can be
distinguished in the puckered structure of a phosphorene
monolayer~\cite{{DT230},{pmet14},{Boulfelfel12}}. The ability of
the different phases to form non-planar in-layer connections at
essentially zero energy cost suggests the possibility to form
faceted nanotube and fullerene structures that are as stable as
planar phosphorene. The possibility to mix different phases within
each wall of spherical and cylindrical single- and multi-wall
structures would offer unprecedented richness not only of form,
but also the associated electronic properties. Bulk quantities of
carbon nanotubes and fibers are currently used as a
performance-enhancing additive to graphite in Li-ion batteries
(LIBs)~\cite{EndoLIB01}. Since black phosphorus is considered
superior to graphite for LIB
applications~\cite{{Lu-patent05},{Park07}}, a similar benefit
could be derived from the presence of phosphorene nanotubes and
related structures.


Here we present a new paradigm in constructing very stable,
faceted nanotube and fullerene structures by laterally joining
nanoribbons or patches of different planar phosphorene phases. Our
{\em ab initio} density functional calculations indicate that
these phases may connect laterally at an angle. Unlike fullerenes
and nanotubes obtained by deforming a single-phase planar
monolayer at substantial energy penalty, we find faceted
fullerenes and nanotubes to be nearly as stable as planar
single-phase monolayers. The resulting rich variety of polymorphs
allows to tune the electronic properties of phosphorene nanotubes
and fullerenes not only by the chiral index, but also by the
combination of different phosphorene phases. In selected PNTs, a
metal-insulator transition may be induced by strain or by changing
the number of walls.

We utilize {\em ab initio} density functional theory (DFT) as
implemented in the \textsc{SIESTA}~\cite{SIESTA} code to obtain
insight into the equilibrium structure, stability and electronic
properties of nanotubes and fullerenes based on different layered
phosphorus allotropes. We use periodic boundary conditions
throughout the study, with nanotubes and fullerenes separated by a
vacuum region exceeding 15~{\AA}. We utilize the
Perdew-Burke-Ernzerhof (PBE)~\cite{PBE} exchange-correlation
functional, norm-conserving Troullier-Martins
pseudopotentials~\cite{Troullier91}, and a double-$\zeta$ basis
including polarization orbitals. Van der Waals interactions are
described using the {\textsc optB86b-vdW}
functional~\cite{{Klimes10},{Klimes11}} as implemented in the
{\textsc VASP}~\cite{VASP} code. We sample the reciprocal space by
a fine grid~\cite{Monkhorst-Pack76} of $8$~$k$-points for 1D
Brillouin zone of nanotubes and only $1$~$k$-point for the small
Brillouin zone of isolated fullerenes. We use a mesh cutoff energy
of $180$~Ry to determine the self-consistent charge density, which
provides us with a precision in total energy of
${\alt}2$~meV/atom. We discuss geometries that have been optimized
using the conjugate gradient method~\cite{CGmethod} until none of
the residual Hellmann-Feynman forces exceeded $10^{-2}$~eV/{\AA}.


\begin{figure*}[t]
\includegraphics[width=1.5\columnwidth]{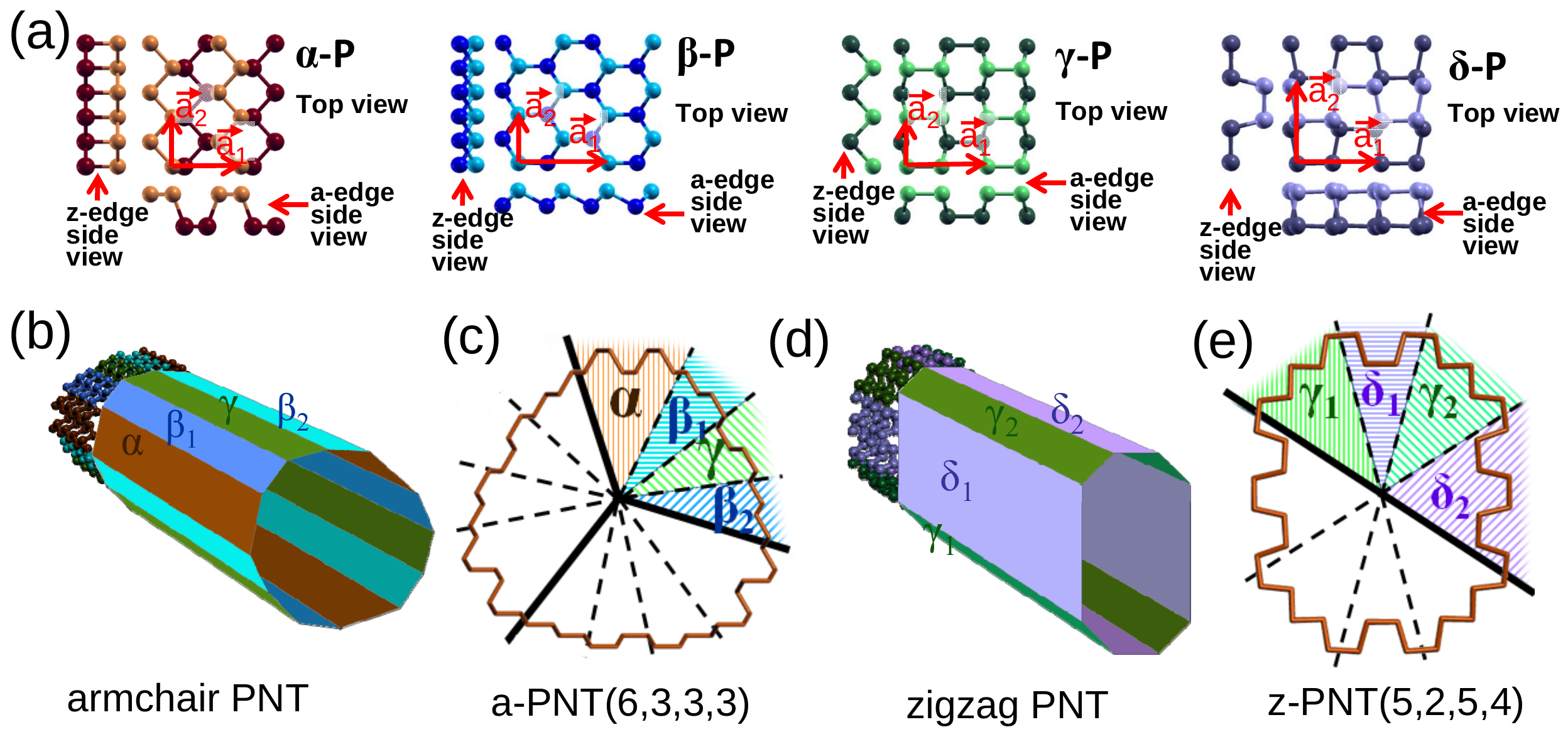}
\caption{(Color online) (a) Atomic structure of $\alpha$-,
$\beta$-, $\gamma$- and $\delta$-P in top view and the side view
of zigzag and armchair edges. The orthogonal lattice vectors
$\vec{a}_1$ and $\vec{a}_2$ define the unit cells or supercells
used in this study. Schematic and atomic structure of (b) an
armchair and (d) a zigzag PNT, with the different structural
phases distinguished by color and shading. The cross-sections of
(c) an armchair and (e) a zigzag nanotube illustrate the symmetry
and the distribution of phases along the perimeter.%
\label{fig1}}
\end{figure*}

The nanotube and fullerene structures presented in this study are
formed by laterally connecting the different stable allotropes of
layered phosphorus, namely $\alpha$-, $\beta$-, $\gamma$- and
$\delta$-P, which are shown in Fig.~\ref{fig1}(a). Whereas
$\alpha$- and $\beta$-P are the most stable allotropes with
$E_{coh}=3.28$~eV/atom in the monolayer, the stability of
$\gamma$- and $\delta$-P is lower only by
$<0.1$~eV/atom~\cite{pmet14}. All these structures share the
underlying honeycomb lattice with graphene, but -- in contrast to
graphene -- are not flat. In analogy to graphene, we define the
armchair and zigzag edges of the different phosphorene phases in
Fig.~\ref{fig1}(a). The vectors $\vec{a}_1$ and $\vec{a}_2$, which
span these lattices, may also be used to identify the edges of
phosphorene nanoribbons (PNRs). Considering the equilibrium
non-planar connections between $\alpha$-, $\beta$-, $\gamma$-P
along zigzag edges and $\gamma$-, $\delta$-P along armchair
edges~\cite{pmet14}, we can design two types of faceted nanotubes.

The exact morphology of the more common carbon nanotubes (CNTs) is
defined by the chiral index $(n_1,n_2)$, which is associated with
the chiral vector $\vec{C}_h=n_1\vec{a}_1+n_2\vec{a}_2$ on a
graphene monolayer. This vector defines the wrapping into a
nanotube and identifies its edge. There is a common distinction
between armchair nanotubes (a-NTs) with an armchair edge and
zigzag nanotubes (z-NTs) with a zigzag edge. A similar convention
could be used when bending monolayers of $\alpha$-, $\beta$-,
$\gamma$- and $\delta$-P to corresponding nanotubes.

The nanotubes we consider here are very different, as they are
formed by connecting planar narrow nanoribbons of different
phosphorene allotropes. Armchair nanotubes (a-PNTs), shown in
Fig.~\ref{fig1}(b) and \ref{fig1}(c), form by connecting laterally
$\alpha$-PNRs with $\beta$- and $\gamma$-PNRs along their zigzag
edges. Virtually no deformation is required to form a nanotube
with $C_3$ symmetry and a polygonal cross-section, shown in
Fig.~\ref{fig1}(c). The three identical $120^\circ$ segments in
the cross-section of this a-PNT contain, in this sequence, an
$\alpha$-PNR connected to a $\beta$-PNR, $\gamma$-PNR, and
$\beta$-PNR. The width of each individual PNR may be zero or
nonzero, giving rise to many different morphologies, illustrated
in the Supplemental Material~\cite{SM-pnt14}. Since the two
$\beta$-PNRs in this segment may also have a different width, we
distinguish them by the subscript. Next, we imagine joining
laterally all nanoribbons of a given phase $\epsilon$ to a wider
ribbon of width $W_{\epsilon}=n_{\epsilon}|\vec{a}_1|$. Obtaining
in this way the values $n_{\alpha}$, $n_{\beta_1}$, $n_{\gamma}$
and $n_{\beta_2}$, we may characterize an armchair nanotube as
a-PNT($n_{\alpha}$,$n_{\beta_1}$,$n_{\gamma}$,$n_{\beta_2}$) and
identify the nanotube in Fig.~\ref{fig1}(c) as a-PNT(6,3,3,3).

In analogy to a-PNTs, zigzag nanotubes (z-PNTs), shown in
Fig.~\ref{fig1}(d) and \ref{fig1}(e), form by connecting laterally
$\gamma$- and $\delta$-PNRs along their armchair edges. Virtually
no deformation is required to form a nanotube with $C_2$ symmetry
and a polygonal cross-section, shown in Fig.~\ref{fig1}(e). The
two identical $180^\circ$ segments in the cross-section of this
z-PNT contain, in this sequence, a $\gamma$-PNR connected to a
$\delta$-PNR, $\gamma$-PNR, and $\delta$-PNR. The width of each
individual PNR may be zero or nonzero, giving rise to many
different morphologies, also illustrated in the Supplemental
Material~\cite{SM-pnt14}. Since the two $\gamma$- and the two
$\delta$-PNRs in this segment may also have a different width, we
distinguish them by the subscript. Next, we imagine joining
laterally all nanoribbons of the same phase $\epsilon$ to a wider
ribbon of width $W_{\epsilon}=n_{\epsilon}|\vec{a}_2|$. Obtaining
in this way the values $n_{\gamma_1}$, $n_{\delta_1}$,
$n_{\gamma_2}$ and $n_{\delta_2}$, we may characterize a zigzag
nanotube as
z-PNT($n_{\gamma_1}$,$n_{\delta_1}$,$n_{\gamma_2}$,$n_{\delta_2}$)
and identify the nanotube in Fig.~\ref{fig1}(e) as z-PNT(5,2,5,4).
We do not discuss here the narrowest z-PNT(1,0,1,0) with a P$_4$
square in the cross-section, which is in reality a nanowire.

Whereas the designation
a-PNT($n_{\alpha}$,$n_{\beta_1}$,$n_{\gamma}$,$n_{\beta_2}$)
defines the way to construct a unique armchair nanotube from PNRs,
a given nanotube may be characterized by different sets of chiral
indices. As discussed in the Supplemental
Material~\cite{SM-pnt14}, this ambiguity stems from the
arbitrariness in assigning atoms at a nanoribbon connection to
either side and can be avoided by selecting $n_{\alpha}=max$. A
similar ambiguity in the nomenclature of z-PNTs can be avoided by
selecting $n_{\gamma_1}=max$ and $n_{\gamma_2}=max$.

%

\begin{figure}[tb]
\includegraphics[width=0.9\columnwidth]{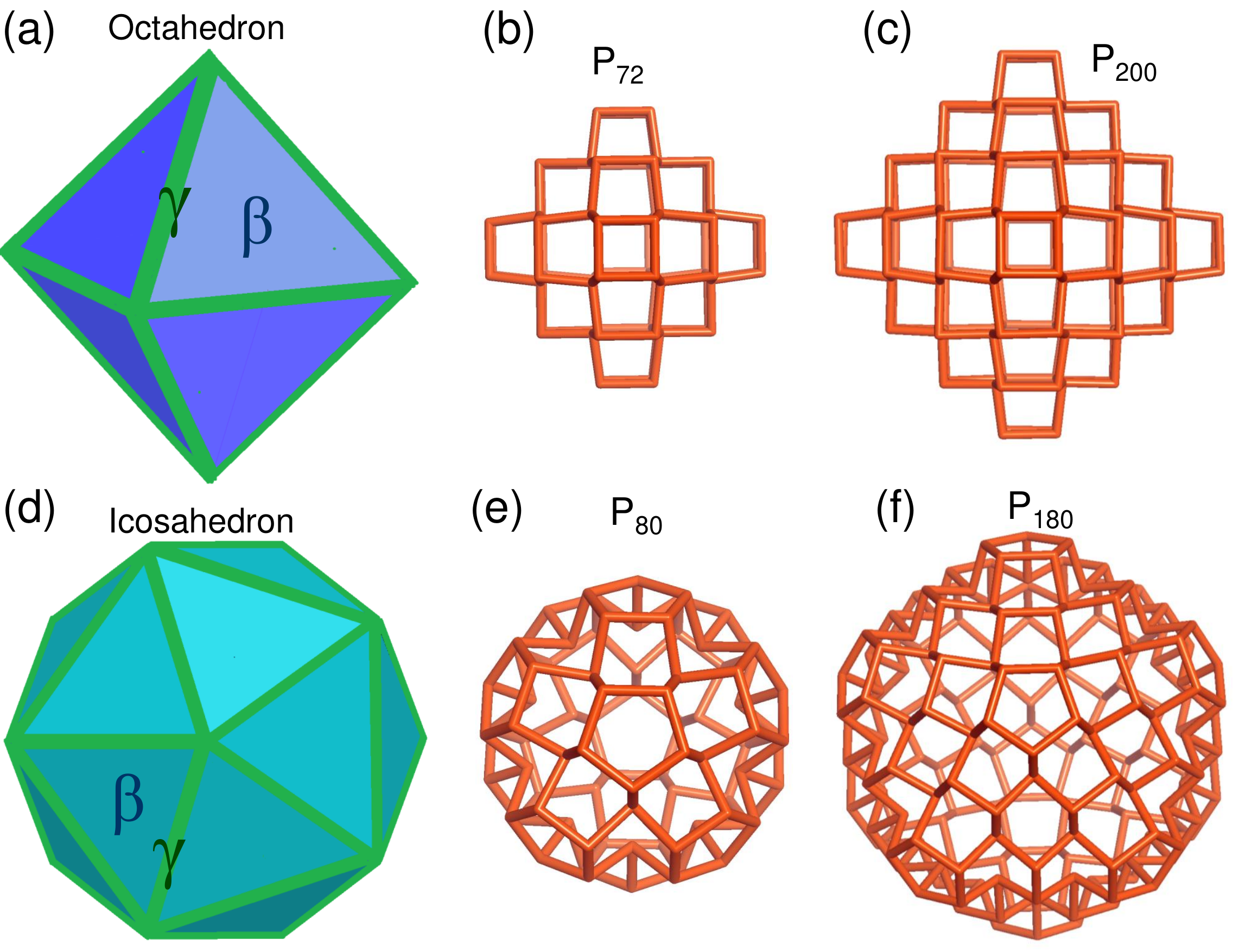}
\caption{(Color online) Phosphorene-based fullerene structures
with (a-c) octahedral and (d-f) icosahedral symmetry. The
structural models in (a) and (d) indicate, how triangular facets
of $\beta$-P are connected by $\gamma$-P along the edges. The
stick models of P$_{72}$ in (b), P$_{200}$ in (c), P$_{80}$ in (e)
and P$_{180}$ in (f) depict the relaxed atomic structures of
octahedral and icosahedral fullerenes.%
\label{fig2}}
\end{figure}

Similar to the construction of nanotubes by connecting nanoribbons
of different phases, also fullerenes may be constructed by
connecting planar triangular segments of $\beta$-P monolayers by
narrow $\gamma$-P strips at the edges, as shown in
Fig.~\ref{fig2}. We have considered octahedral fullerenes,
illustrated schematically in Fig.~\ref{fig2}(a), and icosahedral
fullerenes, illustrated schematically in Fig.~\ref{fig2}(d). Ideal
P$_n$ octahedral fullerenes contain $n=8m^2$ atoms and icosahedral
fullerenes contain $n=20m^2$ atoms, where $m$ is an integer. Two
examples of octahedral fullerenes are presented in
Figs.~\ref{fig2}(b) and \ref{fig2}(c), and two examples of
icosahedral fullerenes in Figs.~\ref{fig2}(e) and \ref{fig2}(f).
Since these structures do not require significant deformation of
the planar monolayer structure, but rather results from an optimum
connection between $\beta$-P and $\gamma$-P, they also are
expected to be nearly as stable as the planar single-phase
allotropes.

\begin{figure}[tb]
\includegraphics[width=1.0\columnwidth]{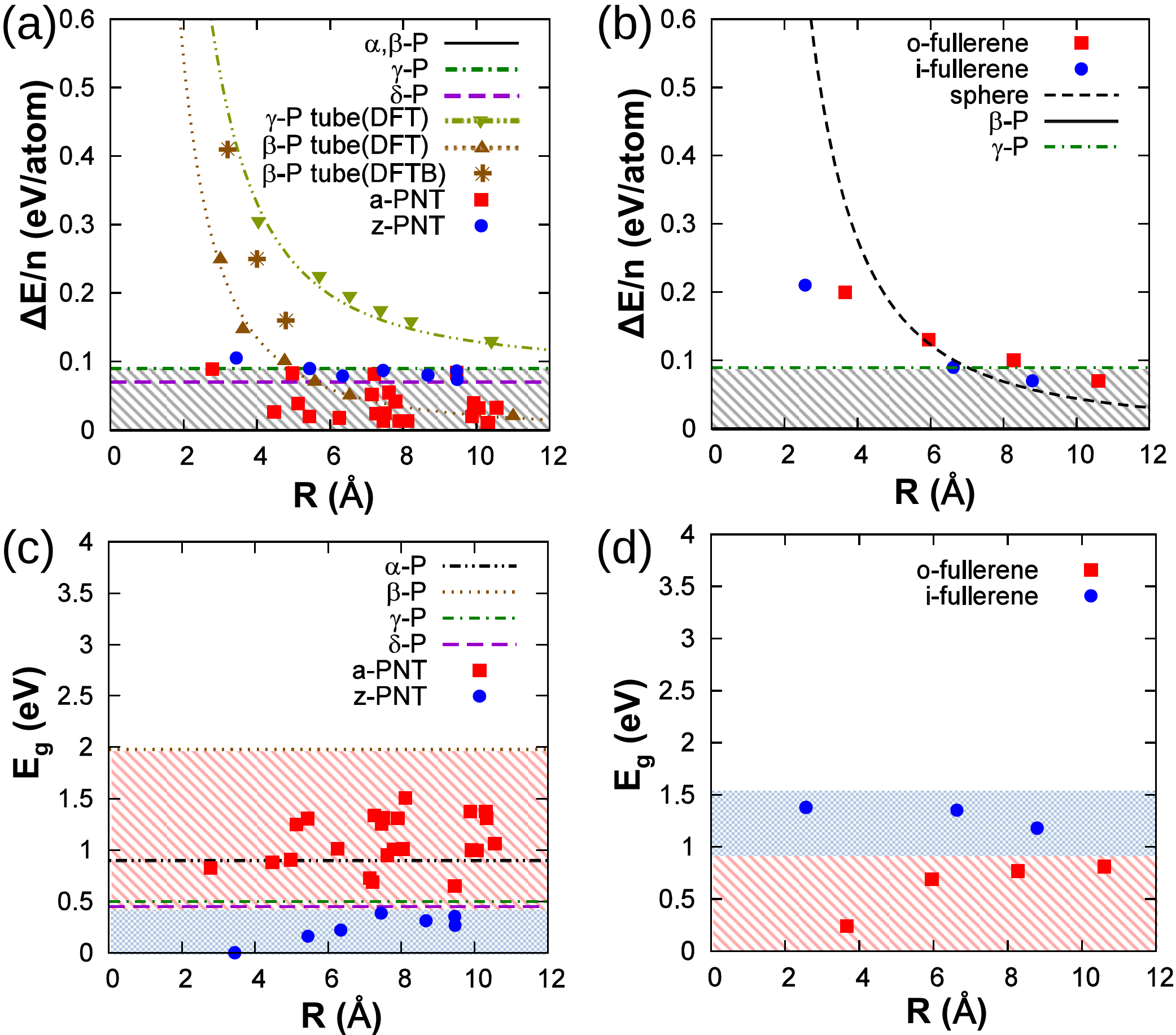}
\caption{(Color online) Stability and electronic structure of
faceted nanotubes and fullerenes. (a) Average strain energy per
atom ${\Delta}E/n$ in PNTs of different radius $R$ with respect to
a planar $\beta$-P monolayer. The shaded region indicates the
range of stabilities of different planar phases and contains most
data points for multi-component faceted nanotubes. For the sake of
comparison, we also present data points for pure-phase PNTs
obtained by rolling up $\beta$- and $\gamma$-P to a tube. (b)
Strain energy per atom in octahedral (o) and icosahedral (i)
fullerenes of radius $R$. The dashed lines in (a) and (b)
represent the $1/R^2$ behavior based on continuum elasticity
theory for pure-phase nanostructures. (c) Fundamental band gaps
$E_g$ in faceted a-PNTs and z-PNTs. The horizontal lines depict
$E_g$ values in pure planar phosphorene monolayers and help to
rationalize the separation between large gap values in a-PNTs and
small gap values in z-PNTs. (d) HOMO-LUMO gaps in o- and
i-fullerenes. O-fullerenes have consistently larger band gaps
than i-fullerenes. %
\label{fig3}}
\end{figure}


Our results for the relative stability of phosphorene nanotubes
are presented in Fig.~\ref{fig3}(a) and those for fullerenes in
Fig.~\ref{fig3}(b). In both sub-figures, the dashed lines display
the expected $1/R^2$ behavior of the strain energy per atom
${\Delta}E/n$ on the radius $R$ that energetically penalizes
structures with small radii.

As seen in Fig.~\ref{fig3}(a), this projected behavior, based on
continuum elasticity theory~\cite{DT071}, agrees closely with our
results for pure $\beta$- and $\gamma$-P nanotubes and previously
published results for $\beta$-P nanotubes, based on density
functional based tight-binding (DFTB)
calculations~\cite{Seifert2000}. As anticipated originally, the
faceted multi-component nanotubes are much more stable than these.
We find that {\em (i)} their strain energies are nearly
independent of the radius and {\em (ii)} their relative
stabilities lie in the value range delimited by the stabilities of
the pure planar components, indicated by the shaded region. Since
z-PNTs contain the least stable $\gamma$ and $\delta$ phases, they
are also least stable among the faceted nanotubes. Presence of the
most stable $\alpha$ and $\beta$ phases, on the other hand, makes
a-PNTs consistently more stable than z-PNTs.

Stability enhancement caused by the coexistence of multiple phases
can also be observed in our results for fullerenes in
Fig.~\ref{fig3}(b). As in the nanotubes, we find most strain
energies within the value range delimited by the pure planar
$\beta$- and $\gamma$-P phases. The stability enhancement is best
visible in very small fullerenes. Interestingly, we find the small
fullerene structures more stable than P$_4$, the building block of
the (most reactive) bulk phosphorus allotrope. Our canonical
molecular dynamics simulations, described in the Supplemental
Material~\cite{SM-pnt14}, show that all nanotube and fullerene
structures we investigated are stable up to $1,000$~K, slightly
above $T_M=863$~K, the melting point of red
phosphorus~\cite{Kittel}.


In carbon nanotubes and fullerenes, the occurrence of a
fundamental band gap is a signature of quantum confinement in the
underlying semi-metallic graphene structure. The advantage of
phosphorene over graphene is the presence of a fundamental band
gap in all layered allotropes discussed here. We thus expect the
fundamental band gaps $E_g$ of nanotubes and fullerenes to
approximately span the value range of the pure components,
indicated by the shaded regions in Figs.~\ref{fig3}(c) and
\ref{fig3}(d). Even though additional corrections are expected due
to quantum confinement and structural relaxation, such corrections
are apparently not as important, since most of our data points lie
in the range delimited by the pure components. At this point, we
wish to point out that our electronic structure results in
Figs.~\ref{fig3}(c) and \ref{fig3}(d), obtained by DFT-PBE, are
expected to underestimate the fundamental band
gaps~\cite{{DT229},{DT230}}.

As seen in Fig.~\ref{fig3}(c), we find larger band gap values in
armchair PNTs containing $\alpha$-, $\beta$- and $\gamma$-P, since
each of the pure planar components has a band gap in excess of
0.5~eV in the monolayer. Since $\gamma$- and $\delta$-P have the
smallest band gaps among the phosphorene allotropes, we also see
the smallest band gaps in z-PNTs, which contain these two pases.
In the narrow z-PNT(3,0,3,0), shown in the Supplemental
Material~\cite{SM-pnt14}, the close distance between third
neighbors along the inner tube perimeter causes the band gap to
close.

In finite-size fullerenes, $E_g$ represents the gap between the
highest occupied molecular orbital (HOMO) and the lowest
unoccupied molecular orbital (LUMO). Our results in
Fig.~\ref{fig3}(d) suggest that the HOMO-LUMO gaps in icosahedral
fullerenes are larger than in octahedral fullerenes. Even though
the values are similar to those of nanotubes in
Fig.~\ref{fig3}(c), we can not easily rationalize the value range
for o- and i-fullerenes, since both structures consist of the same
$\beta$-P and $\gamma$-P allotropes.

\begin{figure}[tb]
\includegraphics[width=1.0\columnwidth]{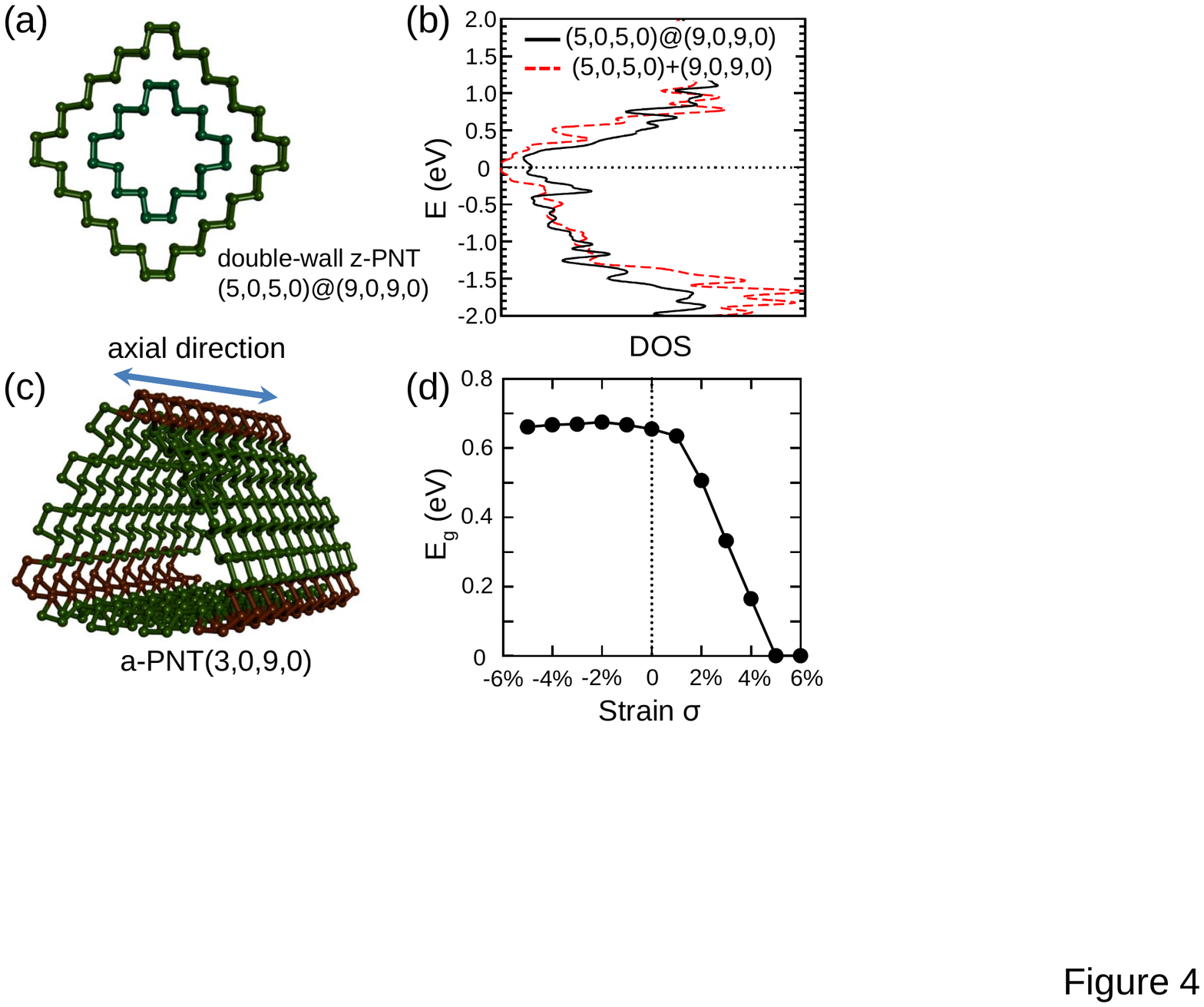}
\caption{(Color online) (a) Cross-section and (b) DOS of the
double-wall z-PNT (5,0,5,0)@(9,0,9,0). The solid line in (b) shows
the total DOS and the dashed line depicts the superposition of the
densities of states of the isolated nanotube components. The Fermi
level $E_F$ is set at 0. (c) Perspective view of the
a-PNT(3,0,9,0) and (d) dependence of the gap energy $E_g$ on axial
strain. \label{fig4}}
\end{figure}

As reported
previously~\cite{{DT229},{pmet14},{DT230},{CastroNeto14},{Yang14}},
the fundamental band gap in phosphorene depends sensitively on the
number of layers and on in-layer strain. Our results in
Fig.~\ref{fig4} indicate that the same behavior occurs also in
PNTs. We find multi-wall PNTs to be stabilized by an inter-wall
interaction of ${\alt}50$~meV/atom, roughly the same as in the
layered
compounds~\cite{{LShulenburger-private},{LShulenburger13}}. Due to
a large fraction of $\gamma$-P in the wall, which has been shown
to undergo a metal-semiconductor transition, we investigated the
(5,0,5,0)@(9,0,9,0) double-wall z-PNT, shown in
Fig.~\ref{fig4}(a). The density of states (DOS) of this
double-wall PNT, shown in Fig.~\ref{fig4}(b), indicates that the
inter-wall interaction may turn two semiconducting nanotubes
metallic upon being combined to a double-wall nanotube. As seen in
Figs.~\ref{fig4}(c) and \ref{fig4}(d) for the single-wall
a-PNT(3,0,9,0), even a modest 5\% stretch may turn a
semiconducting nanotube containing a significant fraction of
$\gamma$-P metallic. This low level of strain may be applied
externally or induced by epitaxy, including structural changes
induced in multi-wall nanotubes.

The most important implication of our claim that faceted nanotubes
and fullerenes are as stable as planar phosphorene is that they
should exist in nature and will be observed eventually, as was the
case with boron nanostructures~\cite{{boron80},{boron40}}. We feel
that phosphorus nanotubes and fullerenes may form during ball
milling of black phosphorus~\cite{Park07} under inert, oxygen-free
atmosphere. This process may also produce structures with a large
accessible surface area for phosphorus-based LIB
applications~\cite{{Lu-patent05},{Park07}}.

In conclusion, we have presented a new paradigm in constructing
very stable, faceted nanotube and fullerene structures by
laterally joining nanoribbons or patches of different planar
phosphorene phases. Our {\em ab initio} density functional
calculations indicate that these phases may form very stable,
non-planar joints. Unlike fullerenes and nanotubes obtained by
deforming a single-phase planar monolayer at substantial energy
penalty, we find faceted fullerenes and nanotubes to be nearly as
stable as the planar single-phase monolayers. The resulting rich
variety of polymorphs allows to tune the electronic properties of
phosphorene nanotubes and fullerenes not only by their chiral
index, but also by the combination of different phosphorene
phases. In selected PNTs, a metal-insulator transition may be
induced by strain or changing the number of walls.

We thank Luke Shulenburger for useful discussions. This study was
supported by the National Science Foundation Cooperative Agreement
\#EEC-0832785, titled ``NSEC: Center for High-rate
Nanomanufacturing''. Computational resources have been provided by
the Michigan State University High Performance Computing Center.



%

\end{document}